\pdfoutput=1
\documentclass{article}
\usepackage{xcolor,colortbl}
\usepackage{spconf,amsmath,graphicx}
\usepackage{lipsum}
\usepackage{bm}
\usepackage{amsmath}
\usepackage{balance}
\usepackage{booktabs}
\usepackage{multirow}
\usepackage{float}
\DeclareMathOperator*{\argmax}{arg\,max}

\usepackage{hyperref}
\usepackage{cleveref}[2012/02/15]

\crefformat{footnote}{#2\footnotemark[#1]#3}

\title{Increase Apparent Public Speaking Fluency By Speech Augmentation}
\name{Sagnik Das \qquad Nisha Gandhi \qquad Tejas Naik \qquad Roy Shilkrot\thanks{Examples: https://sagniklp.github.io/pub-speaker-aug/}}
\address{Human Interaction Lab, Stony Brook University\\
	Department of Computer Science\\
	Stony Brook, NY, USA}
\begin{document}
%\ninept
%
\maketitle
\begin{abstract}
Fluent and confident speech is desirable to every speaker. But professional speech delivering requires a great deal of experience and practice. In this paper, we propose a speech stream manipulation system which can help non-professional speakers to produce fluent, professional-like speech content, in turn contributing towards better listener engagement and comprehension. We propose to achieve this task by manipulating the disfluencies in human speech, like the sounds \textit{uh} and \textit{um}, the filler words and awkward long silences. Given any unrehearsed speech we segment and silence the filled pauses and doctor the duration of imposed silence as well as other long pauses (\textit{disfluent}) by a predictive model learned using professional speech dataset. Finally, we output a audio stream in which speaker sounds more fluent, confident and practiced compared to the original speech he/she recorded. According to our quantitative evaluation, we significantly increase the fluency of speech by reducing rate of pauses and fillers.
\end{abstract}
\section{Introduction}
\label{sec:intro}
Professional speakers, who make their living from their speech, speak clearly and fluently with very few repetitions and revisions. This kind of error-free utterances is the result of many hours of practice and experience. On the other hand, a regular speaker generally speaks with no real practice of articulation and delivery. Naturally, words of an unrehearsed speech contain unintentional \textit{disfluencies} interrupting the flow of the speech. Speech disfluency generally contains long pauses, discourse markers, repeated words, phrases or sentences and fillers or filled pauses like \textit{uh} and \textit{um}. According to the research by \cite{tree1995effects} approximately $6\%$ of the speech appears to be non-pause disfluency. Filled pauses or filler words are the most common disfluency in any unrehearsed, impromptu speech \cite{womack2012disfluencies}. 

Numerous linguistics research has been conducted to find out the effect of speech disfluencies on the listener's comprehension and speakers cognitive state such as uncertainty, confidence, thoughtfulness and cognitive load (\cite{corley2008hesitation, womack2012disfluencies}). Different studies claim that disfluencies often refer to uncertainties in speakers' mind about the future statements. Consequently, less confident speakers tend to be more disfluent.\cite{womack2012disfluencies}. 
\begin{figure}[t]
\centering
\includegraphics[width=0.38\textwidth]{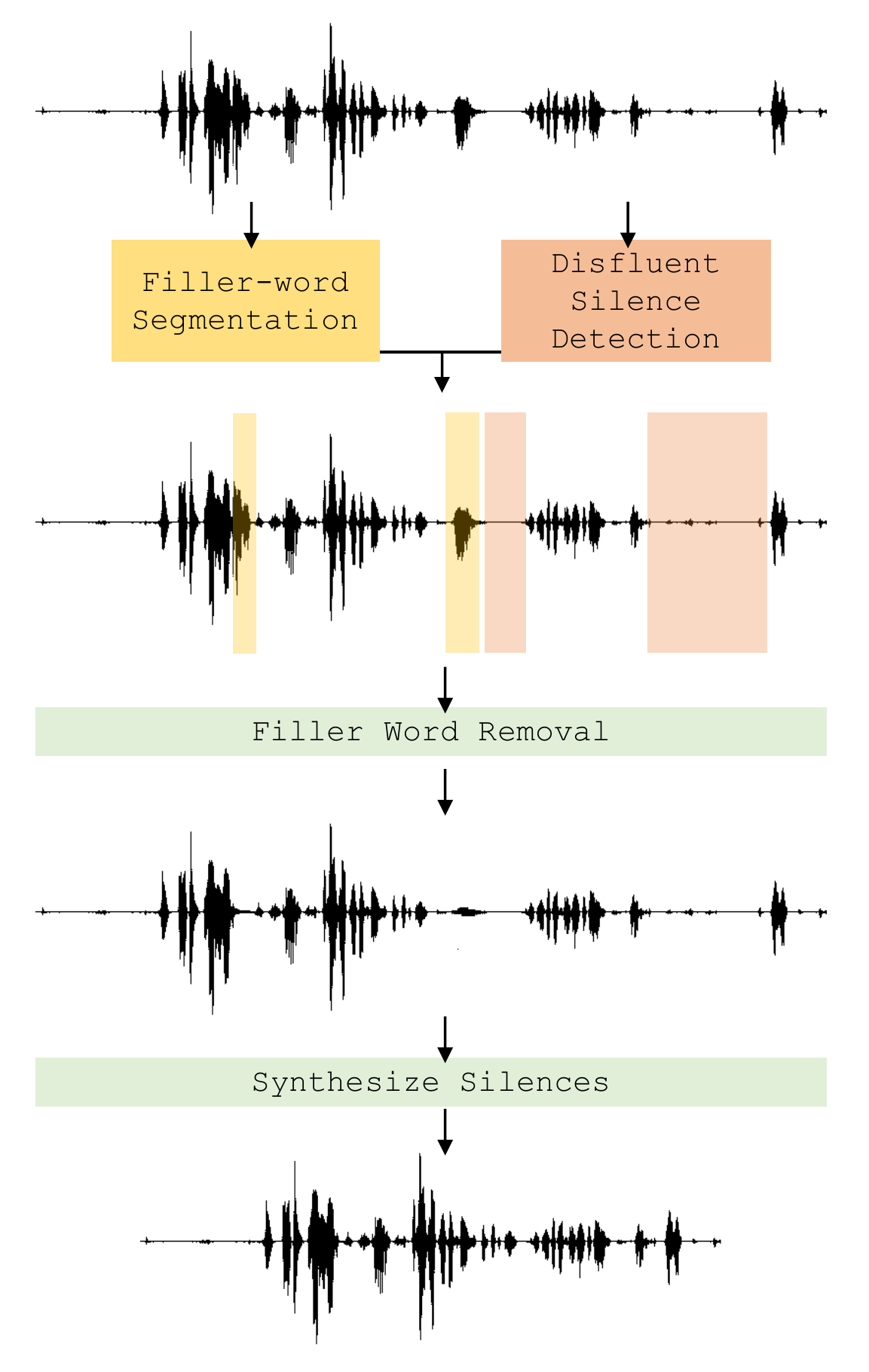}
\label{fig:touchup}
\caption{The proposed speaker augmentation pipeline}
\end{figure}
Moreover, it has also been observed that filled pauses specifically indicate the level of cognitive difficulty of the speaker \cite{barr2010role}. Generally, disfluencies occur before a longer utterance \cite{shriberg1996disfluencies} or when the topic is unfamiliar to the speaker \cite{merlo2004descriptive}. Fluency reflects the speaker's ability to focus the listener's attention on his/her message rather than inviting the listener to focus on the idea and try to self-interpret it \cite{lennon1990investigating}. Considering the diverse factors affecting speaker fluency, our idea is to doctor a speech to make it fluent by taking care of temporal factors contributing to it.

In this work, we propose a system to detect, segment, and remove the most common disfluencies, namely the filler words and long, unnatural pauses from a speech to aid speakers' fluency. Our system takes a raw speech track as input and outputs a modified fluent version of it by intelligently removing the filled pauses and adjusting the ``long" silences. We evaluate the performance of our system quantitatively on speeches of non-native speakers of English. We also propose an assistive user interface which can be used to help users' to visualize and comparatively analyze their speech.   

We interpret the occurrence of disfluencies in a speech as an acoustic event, and a segmentation approach is taken for the detection. A CNN and RNN combined architecture, a convolution recurrent neural network (CRNN) is used to achieve the task, inspired from \cite{cakir2017convolutional}. Further, a binary classification approach is taken to detect long pauses between words. After deleting the filler-words and adjusting the silences the fluent version of the speech is obtained. The performance of our system is evaluated on speeches of non-native speakers of English using fluency metrics proposed by \cite{kormos2004exploring}. We also propose an assistive user interface which can be used to help users' to visualize and comparatively analyze their speech. The essential contributions of this paper are -
\begin{enumerate}
\item A disfluency detection mechanism that works directly on acoustic features without using any language features.
\item A silence modeling scheme directly conditioned on the previous speech.
\item A disfluency repair technique to help users improve a pre-delivered speech.

\end{enumerate}
\section{Related Works}
\label{sec:related}

In recent years, there have been many works related to speech disfluencies, spanned across the domains of psychology, linguistics and natural language processing (NLP). Where, the psychology and linguistic researchers focused on defining disfluencies, the reasons and effects of it from a language and cognitive aspect; the NLP researchers focused more on detecting these from speech transcripts to help language understanding and recognition systems.

% \subsection{Linguistic and Psychological aspects of Disfluencies}
% In psycholinguistics, \textit{disfluencies}, more specifically, \textit{filler-words} is a long studied subject.
The prime motivation for disfluency detection in NLP is to better interpret the speech-to-text transcripts for natural language understanding systems.

One of the first work \cite{charniak2001edit}, focuses on classifying the edit words (restarts and repairs) from the text using a boosted classifier. Another contemporary method was to apply a noisy channel model to detect and correct speech disfluencies \cite{honal2003correction,johnson2004tag,zwarts2010detecting}. Later, Hidden Markov Model (HMM), Conditional Random Field (CRF), Integer Linear Programming (ILP) based \cite{liu2006enriching,georgila2009using} methods are introduced. A classification approach using lexical features is taken by \cite{howes2012helping}, they specifically focus on schizophrenic patient dialogs. Some incremental \cite{honnibal2014joint,hough2014strongly,howes2014helping,ferguson2015disfluency}, multi-step \cite{qian2013disfluency} and joint task (parsing and disfluency detection) \cite{rasooli2013joint,honnibal2014joint} methods were introduced recently. Though all these provide some convincing results, all of them are limited to pre-defined feature templates (lexical, acoustic, and prosodic). 

With advances in deep learning, most recent methods rely on recurrent neural networks (RNN) \cite{hough2015recurrent,wang2016neural,zayats2016disfluency,hough2017joint,wang2017transition}. These methods use word embeddings and acoustic features instead of pre-defined feature templates. 

All the techniques above, make one fundamental assumption, i.e., any disfluency detection must have an automatic speech recognizer (ASR) in the pipeline. Consequently, to the best of our knowledge till now all the presented disfluency detection schemes work in the transcript level. Also, these systems have never been paired with an acoustic level repair scheme with a goal of exploring the use-cases from the perspective of the listener and the speaker.  

In our work, we address these motivations by devising a disfluency detection and repair method relying solely on acoustic features to synthesize temporally fluent speech segments from the perspective of human-interaction. 
\section{Proposed Method}
\label{sec:pagestyle}
\begin{figure*}[t]
\centering
\includegraphics[width=0.9\textwidth]{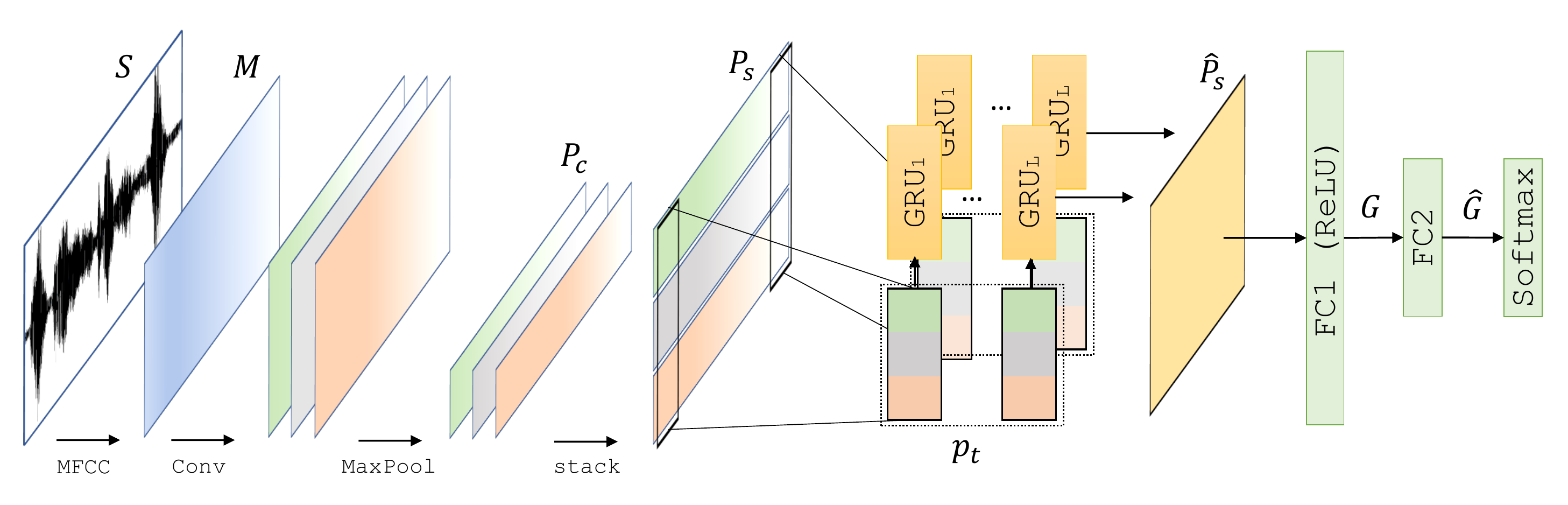}
\label{fig:block}
\caption{Block diagram of the filler-word segmentation}
\end{figure*}
\subsection{Disfluency Detection}
Our work focuses on building a system that can be used not only as a disfluency detection system but also provide a way to understand users' disfluency better. The primary motivations of this work are the following-
\begin{itemize}
\item Work with disfluencies on the acoustic level without using any transcript. 
\item Significant portion of a disfluent speech contains long pauses. In transcript level, it's not an issue, but in the acoustic level, it matters a lot in determining speakers' fluency.
\item Repairing disfluent segments to help users understand the possible improvements to their speech, as well as allow to create fluent speech content without much hassle. 
\end{itemize}
The types of disfluencies we considered in this work, are the use of filler words, and intermittent long pauses.  
\subsubsection{Dataset}
The dataset used for filler word segmentation is obtained from Switchboard transcription\footnote{https://www.isip.piconepress.com/projects/switchboard/}. We also used the Automanner\footnote{https://www.cs.rochester.edu/hci/currentprojects.php?proj=automanner} \cite{tanveer2016automanner} transcription for additional data. This gives more generalization to our training samples since contains recording from standard interfaces.    
% There are standard datasets labeled for disfluency detection but, labelings are made at the token level of transcripts. To work on acoustic level accurate time windows for all the filler words and disfluent silences were needed. We used existing disfluency detection method and ASR output to label speech onset/offset timeline for disfluencies. 

To label disfluent silences we use combination of a silence probability model \cite{chen2015pronunciation} and a disfluency detection model \cite{zayats2016disfluency}. First, we locate the silences and segment each word pair from the dataset then according to the probability model it's decided if silence is disfluent. For each word pair utterance the silence probability model gives a probability of a silence ($P_{sil}$) occurring between them. A word pair with low $P_{sil}$ but a significant amount of silence is labeled as disfluent. If a word pair doesn't exist the model vocabulary, we resort to the following approach. Since, general disfluencies accompany longer silences, any silence within a disfluent segment is labeled as an unnatural pause. Additionally, the word pairs surrounded with silences more than $0.7$ seconds are also labeled similarly. This choice is experimental and can be considered safe because it's considerably higher than the suggested quantitative measure of micro-pauses (\textit{fluent}), 0.2 secs. \cite{riggenbach1991toward}. On the other hand, additional fluent pairs are collected from TIMIT \cite{garofolo1993darpa}.

\subsubsection{Features}
In this step, frame level acoustic features (log mel band energy or mel frequency cepstral coefficients (MFCCs)) are obtained at each timestep $t$ resulting a feature vector $\bm m_t \in \mathcal{R}^C$. Here, $C$ is the number of features (in frequency dimension) at frame $t$. The task of segmenting the filler words is formulated as binary classification of each frame to its correct class $k$ (Eq. \ref{eq:1}).
\begin{equation}
\argmax_{k} ~ P(y_t^{(k)}~|~\bm m_t,\bm\theta)
\label{eq:1}
\end{equation}
Where, $k = \{1,2\}$ and $\bm\theta$ are the parameters of the classifier. In the training data, the target class $~y_t^{(k)}=1$ if frame $t$ belongs to class $k$ (determined using the onset/offset timeline of $k$ associated with a sound segment) , otherwise zero.

Each soundtrack $S$, is divided into multiple fixed length sequences of frames $\bm M_{t:t+T-1}$. Where, $T$ is the length of the frame sequence. The corresponding class label matrix, $\bm Y_{t:t+T-1}$ contains all the $y_t$. 
% We will denote $\bm M_{t:t+T-1}$ as $\bm M$ and $\bm Y_{t:t+T-1}$ as $\bm Y$ throughout the paper for the ease of representation. 
\subsubsection{CRNN for filler word segmentation }
Here we propose a Convolutional Recurrent Neural Network (CRNN) for filler word segmentation. Similar, architecture is previously used for sound event detection (SED) \cite{cakir2017convolutional} and speech-recognition \cite{sainath2015convolutional} task. The architecture is a combination of convolutional and recurrent layers, followed by feed-forward layers.

The sequence of extracted features $\bm M \in \mathcal{R}^{C\times T} $ is fed to the CNN layers with rectified linear unit (ReLU) activations. Filters used in the CNN layers are spanned across the feature and time dimension. Max-pooling is only applied over the frequency dimension. Output of max-pooling is a tensor $\mathcal{P}_c \in \mathcal{R}^{F \times M' \times T}$. Where, $F$ is the number of filters of the final convolution layer, $M'$ is the truncated frequency dimension after the max-pooling operation.

To learn the features over time axis, $F$ feature maps are then stacked along the frequency axis which outputs a tensor $\mathcal{P}_s \in \mathcal{R}^{(F \times M')\times T}$. This is fed to the RNN as a sequence of frames $\bm p_t$ which outputs a hidden vector $\hat{\bm p}_t$. The $i$th recurrent layer output is given as in Eq. \ref{eq:2}. Where, $\mathcal{F}$ is a function learned by the each RNN unit. In this work, we use GRUs as presented in \cite{cho2014learning}.
\begin{equation}
\hat{\bm p}_t^i = \mathcal{F}(\hat{\bm p}_t^{i-1},\hat{\bm p}_{t-1}^i)
\label{eq:2}
\end{equation}

RNN final layer outputs $\hat{\bm p}_t^f$ are then fed to a fully-connected (FC) layer with ReLU activation and $\mathcal{G} \in \mathcal{R}^{FC_1 \times T}$ is obtained where, $FC_1$ is the number of neurons of the layer. Finally, another layer with softmax activation is applied to get the class probabilities. Let, $\hat{\mathcal{G}} \in \mathcal{R}^{K \times T}$ is the output tensor of the final FC layer, then probabilities are given by-
\begin{equation}
P(\bm y_t~|~\bm m_{0:t},\bm\theta)= Softmax(\hat{\bm g}_t)
\end{equation}
The CRNN training objective is to minimize the cross-entropy loss with $l_2$ regularization (Eq. \ref{eq:loss})
\begin{equation}
L(\bm\theta)=-\sum_{0:t} \sum_k \log P(y_t^{(k)}) + \lambda ||\bm\theta||
\label{eq:loss}
\end{equation}

\begin{figure}[t]
\centering
\includegraphics[width=0.48\textwidth]{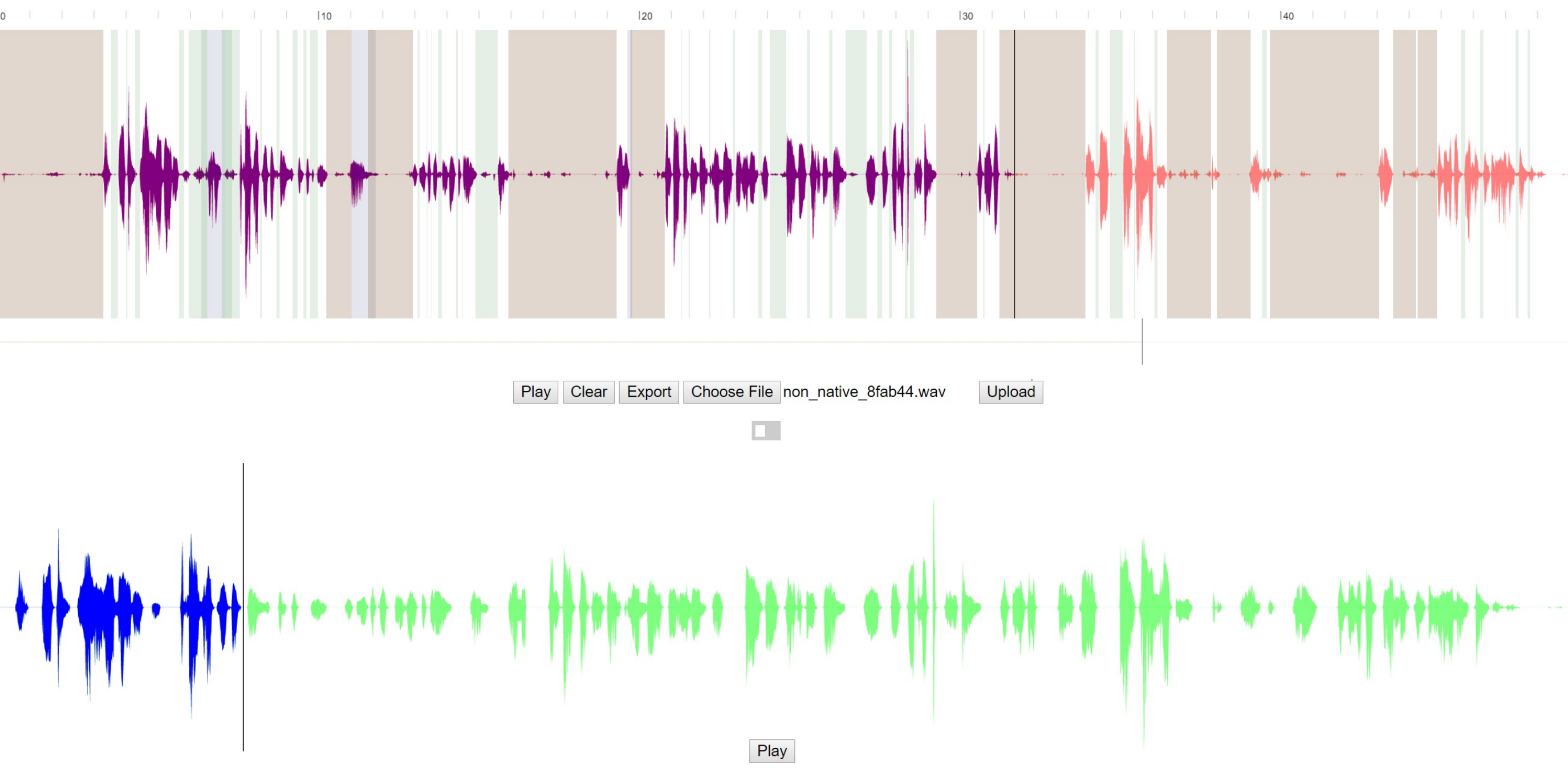}
\caption{\textbf{The visualization interface}; \textit{top}: Speech track with colored segmentation outputs (\textit{brown}: disfluent silences, \textit{blue}: fillers, \textit{green}: fluent silences); \textit{bottom}: Modified speech.}
\label{fig:vis}
\end{figure}

\subsubsection{Disfluent silence Classification}
The problem is formulated as a binary classification task, given a silent segment $Z$, the task is to decide whether it's a disfluent or a non-disfluent silence. Classifying a silence only makes sense when it's combined with adjacent utterances. Because an occurrence of silence is solely driven by the utterance and also heavily influenced by disfluencies. Thus, it's not always evident that all pauses higher than a significant threshold is disfluent, illustration in Fig.\ref{fig:vis} gives an idea of the fact.

We train a support vector machine (SVM) to achieve this task. Given a silent segment Z, it's first padded with the one-word utterances on the left and right ($\hat{Z}$). Then, the MFCC features are extracted and we take the mean over the frequency axis to create the feature vector $\bm z_i \in \mathcal{R}^T$. $T$ is the number of frames in $\bm z_i$. Segments are of variable length thus $\bm z_i$ padded with trailing zeros prior the classification. During, testing we don't use the previous and next word boundaries but a fixed length time window is used. In our experiments, we found that $0.8-1.0$ secs. give pretty good results.

\subsection{Disfluency Repair}
First, the fillers are replaced with silences. We found that it's often helpful (such as when ambient noise is present) to use a decomposition mechanism \cite{rafii2012music} on the speech to separate the background noise and vocals. Then, the fillers are replaced with its corresponding background segment. The modified track is then used to segment ($Z$) the silences and finally, the classification is done.

\begin{figure}[t]
\centering
\includegraphics[width=0.46\textwidth, height=1.7in]{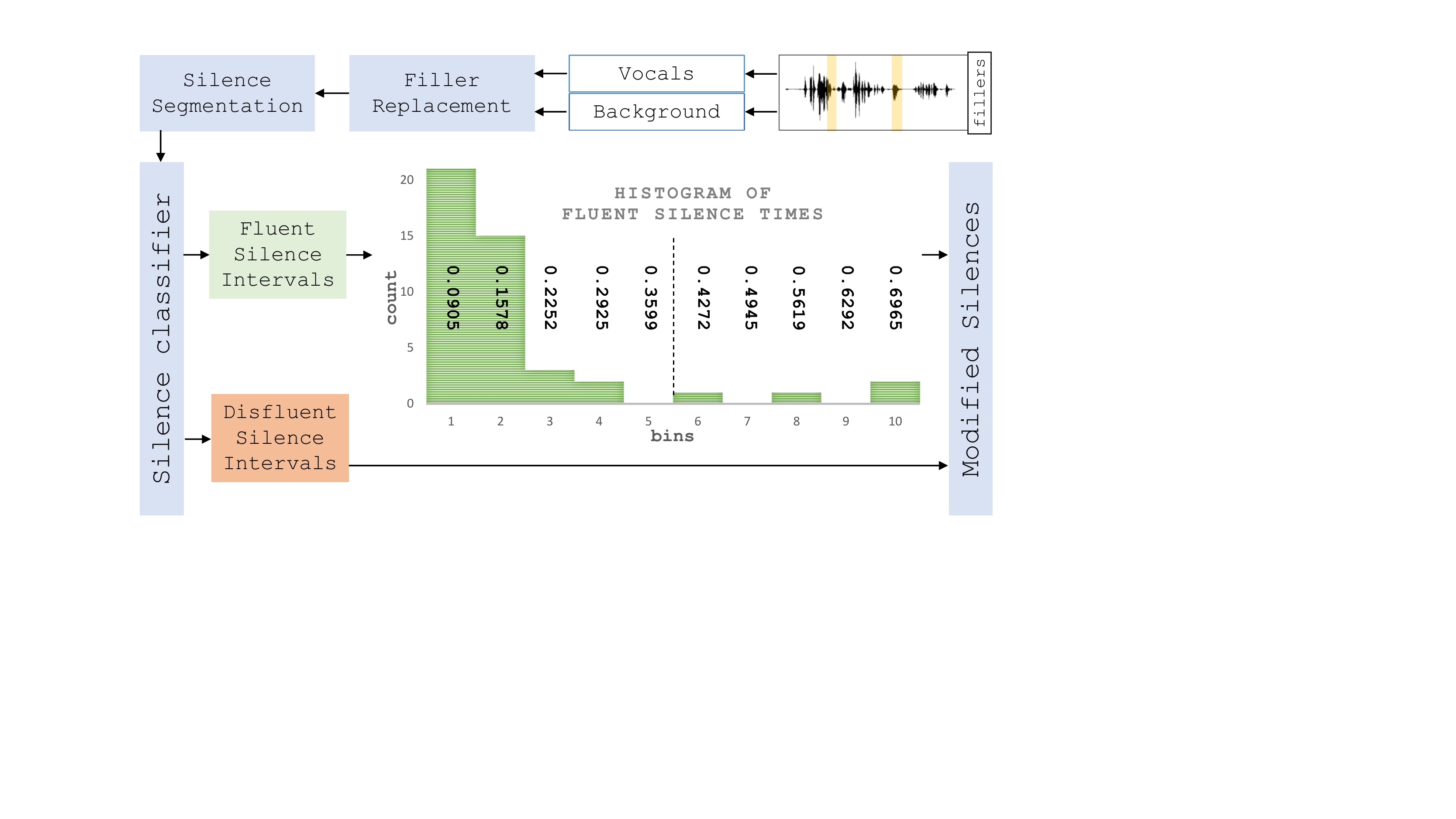}
\caption{Silence modification pipeline: The dashed line on the histogram shows the median time of the fluent silences.}
\label{fig:sil}
\end{figure}

All the silence segment lengths are then modified to make the speech fluent (Fig. \ref{fig:sil}). The goal is to reduce the amount of long, unnatural pauses that hurt the fluency of the speech. It is also required to keep the pace of the speech intact. Too much reduction of silences makes the speech unnatural and broken. We take the fluent silence times (i.e., as suggested by our silence classifier) and obtain a histogram and found that taking the median of the histogram bins as the optimal amount of silence works quite well. In this way, the distribution of the silence along the speech progression confines to a constant distribution and speaker sounds more consistent and fluent in the modified speech. 

\section{Results \& Analysis}
\subsection{Experimental Settings}
\subsubsection{Datasets}
The experiments are performed on Switchboard \cite{godfrey1992switchboard}, Automanner  \cite{tanveer2016automanner} and our dataset of public speaker recording. To train the CRNN we use the segments from the Switchboard. The CRNN test results are reported on held-out data from Switchboard-I. Silence classification results are reported on TIMIT \cite{garofolo1993darpa}, Switchboard, and Automanner held-out dataset. All the fluency metrics are evaluated on our dataset, containing recordings of 20 non-native speakers of English. The speakers were asked to talk on a specific topic for 50-60 seconds.     

\subsubsection{Parameter settings}
We experimented with different configurations of the CNN and RNN parameters and different features.
% \begin{table}[h!]
% \centering
% \begin{tabular}{@{}ccc@{}}
% \toprule
% \textbf{Task} & \textit{mfcc} & \textit{log mel}  \\ \midrule
% Filler Seg. & $40 \times t$ & $128 \times t$  \\
% Sil. Classif. & $40 \times t$ & -  \\ \bottomrule
% \end{tabular}
% \caption{Final feature dimensions used in this work; $t$ is the number of timesteps}
% \label{tab:feats}
% \end{table}

\textit{Types of features}: Initial experiments were performed on \textit{mel frequency cepstrum coefficients} (\textit{mfcc}), mel spectrograms, log mel spectrograms (\textit{log mel}), spectral contrast, zero crossing rate and tonnetz. According to the experimental results, the \textit{mfcc} ($40 \times t$) and \textit{log mel} ($128 \times t$) features are used for filler segmentation. For the silence classification \textit{mfcc} features are used, after taking mean over the frequency axis. The used feature dimensions are shown in table. All features are extracted in 30ms frames with 15ms overlap.

\textit{RNN \& CRNN parameters}:
In experiments with the CNN and CRNN, we explore $\{1,2,3\}$ convolutional (\textit{conv}) layers with combination of max pooling and average pooling. At each layer, ReLU activation is used. Following settings are used for \textit{conv} filters- $\{16,32,64\}$ and kernel sizes- $\{2,3,4,5,8\}$. All the \textit{conv} layers use \textit{same} padding. Pooling size was varied within $\{2,3,4,5,8\}$. The pooling is performed only on the frequency dimension. We tried different dropout ratios of $\{0.3,0.5,0.75\}$. 

The RNN we use is Gated Recurrent Units (GRU). Experiments are performed with $\{2,3\}$ layers (l) and $\{64,128,256\}$ hidden units (d). No intermediate dropout (dr) is applied. 

Final fully connected layer (FC1 in Fig.\ref{fig:block}) is experimented with hidden units (d) of $\{100,200\}$ with dropout ratios of $\{0.3,0.5,0.75\}$. 

\begin{table}[h!]
\centering
\resizebox{0.48\textwidth}{!}{%
\begin{tabular}{@{}cccc@{}}
\toprule
\textbf{Features} & \textbf{CNN} & \textbf{RNN} & \textbf{FC} \\ \midrule
\textit{mfcc} & \begin{tabular}[c]{@{}c@{}}conv1 {[}32,(8,8){]}, conv2 {[}64,(4,4){]}\\ maxpool1 {[}5,5{]}, maxpool2{[}4,4{]}\\ dr=0.25\end{tabular} & \multirow{2}{*}{\begin{tabular}[c]{@{}c@{}}l=3\\ d=128\end{tabular}} & \multirow{2}{*}{\begin{tabular}[c]{@{}c@{}}d=100\\ dr=0.5\end{tabular}} \\
\textit{log mel} & \begin{tabular}[c]{@{}c@{}}conv1 {[}32,(8,8){]}, conv2 {[}64,(4,4){]}\\ maxpool1 {[}8,4{]}, maxpool2{[}4,2{]}\\ dr=0.25\end{tabular} &  &  \\ \bottomrule
\end{tabular}%
}
\caption{Final parameters for CNN, RNN and fully connected layers}
\label{tab:params}
\end{table}
The networks are trained in an end-to-end fashion using AdaGrad algorithm for 200 epochs. The learning rate was set to $0.01$. The regularization constant ($\lambda$) was set to $0.01$. The final parameters are given in Table \ref{tab:params}.

\begin{table*}
\centering
\begin{tabular}{@{}cccccccc@{}}
\toprule
\textbf{Metrics $\rightarrow$}    & \textit{\textbf{SR $\uparrow$}} & \textit{\textbf{AR $\uparrow$}} & \textbf{PTR $\uparrow$} & \textbf{MLR $\uparrow$} & \textbf{MLP $\downarrow$} & \textbf{FPM $\downarrow$} \\ \midrule
\textbf{Original}  & 165.3571             & 171.0986             & 58.865       & 0.400               & 0.654        & 3.659        \\
\textbf{Processed} & \textbf{186.241}              & \textbf{186.241}              & \textbf{65.570}       & \textbf{0.495}           & \textbf{0.365}        & \textbf{1.762}
\\ \bottomrule
\end{tabular}
\caption{The fluency metrics, before and after processing the speeches. $\uparrow$ means higher is better and $\downarrow$ denotes lower is better}
\label{tab:fluencymet}
\end{table*}
\textit{Silence classification parameters}: Max length of the sequences were set to 128. Final parameters are given in table \ref{tab:silclsfparams}.
\begin{table}[H]
\centering
\begin{tabular}{@{}ccc@{}}
\toprule
\textbf{SVM} & \textbf{LogReg} & \textbf{XGBoost} \\ \midrule
\begin{tabular}[c]{@{}c@{}}itr=1500\\ kernel=$rbf$\\ C=10\end{tabular} & \begin{tabular}[c]{@{}c@{}}itr=100\\ C=10\end{tabular} & \begin{tabular}[c]{@{}c@{}}depth=3\\ lr=0.1\\ estimators=100\end{tabular} \\ \bottomrule
\end{tabular}
\caption{Final parameters used in silence classification}
\label{tab:silclsfparams}
\end{table}

\subsubsection{Evaluation Metrics}
\label{sec:metric}
To evaluate the filler word segmentation we use the following frame level statistics:
\begin{itemize}
\item $F_1$ Score ($F_1$): The $F_1$ score is calculated on frame level (30ms) using the TP, the frames where fillers are correctly detected; TN, the frames where non-fillers are correctly detected; FP, the frames where fillers are wrongly detected; and FN, the frames where non-fillers are wrongly detected.
\end{itemize}
The silence classification is evaluated using the $F_1$ score w.r.t. the disfluent silence class.

To evaluate the quality of the augmented speech from our system, we use the following metrics defined in \cite{kormos2004exploring}:
\begin{itemize}
\item \textit{Speech rate}: Is obtained as-
\begin{equation}
SR=\frac{\# \ of\ syllables}{total\ time - ufp [<3]} \times 60
\end{equation}
Where, $ufp [<3]=$ total time of unfilled pauses lesser than 3 seconds. Since, pauses $>3$ secs. are considered as articulation pauses \cite{riggenbach1991toward}. 
\item Articulation rate
\begin{equation}
AR=\frac{\# \ of\ syllables}{total\ time} \times 60
\end{equation}
\item Phonation-time ratio
\begin{equation}
PTR=\frac{speaking\ time}{total\ time}
\end{equation}
\item Mean length of runs
\begin{equation}
MLR=\frac{\# \ of\ syllables}{\#\ utterances\ between\ p[>0.25]}
\end{equation}
Where, $p [>0.25]=$ pauses greater than 0.25 seconds.
% \item Number of long silent-pauses per min.
% \begin{equation}
% PPM=\frac{\# \ of\ p[>0.2]}{speaking\ time} \times 60
% \end{equation}
\item Mean length of pauses
\begin{equation}
MLP=\frac{total\ of\ p[>0.2]}{\# \ of\ p[>0.2]}
\end{equation}
\item Filled pauses per min.
\begin{equation}
FPM=\frac{\#\ of\ filled\ \-pauses}{total\ time}
\end{equation}
\end{itemize}

\subsection{Filler Word Segmentation}
The filler word segmentation performance is evaluation results are given in Table \ref{tab:perf} and \ref{tab:asrvcrnn}. In Table \ref{tab:perf} we report the comparative performance of the CRNN using different features. To understand more about the credibility of the CRNN, in Table \ref{tab:asrvcrnn} we show the results compared to an automatic speech recognizer available with Kaldi (ASpIRE Chain Model\footnote{https://github.com/kaldi-asr/kaldi/tree/master/egs/aspire}). Considering the simplicity of our network, it performs pretty close to the ASR in terms of $F_1$ score. All results are evaluated on a subset of Switchboard-I dataset.
\begin{table}[H]
\centering
\begin{tabular}{@{}cccc@{}}
\toprule
\textbf{Features} & \textit{Precision} & \textit{Recall} & $F_1$ \\ \midrule
\textit{mfcc} & 0.9482 & 0.9610 & 0.9534 \\
\textit{log mel} & \textbf{0.9495} & \textbf{0.9629} & \textbf{0.9550} \\ \bottomrule
\end{tabular}%
\caption{Performance of the CRNN with different features}
\label{tab:perf}
\end{table}
% \begin{table*}[h!]
% \centering
% \begin{tabular}{c|cc|cc|cc|}
% \cline{2-7}
% \textbf{}                             & \multicolumn{2}{c|}{$Precision$} & \multicolumn{2}{c|}{$Recall$}  & \multicolumn{2}{c|}{\textbf{$F_1$}}                       \\ \hline
% \multicolumn{1}{|c|}{\textbf{Method}} & \textit{mfcc}                 & \textit{log mel}    & \multicolumn{1}{c}{\textit{mfcc}} & \multicolumn{1}{c|}{\textit{log mel}} & \multicolumn{1}{c}{\textit{mfcc}} & \multicolumn{1}{c|}{\textit{log mel}} \\ \hline
% % \multicolumn{1}{|c|}{\textbf{CNN}}    &                               &                     &                                   &                      &                &                 \\
% % \multicolumn{1}{|c|}{\textbf{RNN}}    &                               &                     &                                   &                        &         &               \\
% \multicolumn{1}{|l|}{\textbf{CRNN}}   & 0.9482    &         0.9495           &      0.9610                            &        0.9629         &       0.9534      &    0.9550                 \\ \hline
% \end{tabular}
% \caption{Performance of filler word segmentation architectures}
% \label{tab:perf}
% \end{table*}

\begin{table}[H]
\centering
\begin{tabular}{@{}cccc@{}}
\toprule
\textbf{Method} & \textit{$Precision$} & \multicolumn{1}{c}{\textit{$Recall$}} & $F_{1}$ \\ \midrule
\textbf{ASR}    &        \textbf{0.9774}              &             \textbf{0.9792}                          &    \textbf{0.9775}     \\
\textbf{CRNN}   &        0.9495              &              0.9629                         &    0.9550     \\ \bottomrule
\end{tabular}
\caption{Performance of filler word segmentation compared to an automatic speech recognizer.}
\label{tab:asrvcrnn}
\end{table}
The only drawback that we have observed while comparing our method and ASR is that, sometimes our classifier detects segments that sounds similar with 'uh' or 'um'.  
\subsection{Disfluent Silence Classification}
For this task we experimented with SVM, Logistic Regression (LogReg) and XGBoost. The results are summarized in table \ref{tab:silclsf}. We used 10-fold cross validation to report our results.
\begin{table}[H]
\centering
\begin{tabular}{@{}cccc@{}}
\toprule
\textbf{Method $\rightarrow$} & \textbf{SVM} & \textbf{LogReg} & \textbf{XGBoost} \\ \midrule
$F_1$ & 0.9055 & 0.9200 & \textbf{0.9207} \\ \bottomrule
\end{tabular}
\caption{Silence classification performance on TIMIT, SwitchBoard and Automanner}
\label{tab:silclsf}
\end{table}
\subsection{Disfluency Repair}
After processing the speeches by removing the fillers and long silences, the fluent speech is obtained. To compare the fluency of the synthesized and the original speech, discussed metrics (Section \ref{sec:metric}) are used. The results are reported in table \ref{tab:fluencymet}. Mean of each metric across all the samples are reported. From the numbers, it's pretty clear that we improve the fluency. It's notable that in the processed speech the articulation and speech rate increases to same quantity since we take care of all the unfilled pauses in the speech and introduce a more uniform silence production. Apart from the numbers, for qualitative understanding, some processed samples are available \href{https://sagniklp.github.io/disfluency-removal-api/}{\textbf{here}}.

\label{sec:typestyle}

% To achieve the best rendering both in printed proceedings and electronic proceedings, we
% strongly encourage you to use Times-Roman font.  In addition, this will give
% the proceedings a more uniform look.  Use a font that is no smaller than nine
% point type throughout the paper, including figure captions.

% In nine point type font, capital letters are 2 mm high.  {\bf If you use the
% smallest point size, there should be no more than 3.2 lines/cm (8 lines/inch)
% vertically.}  This is a minimum spacing; 2.75 lines/cm (7 lines/inch) will make
% the paper much more readable.  Larger type sizes require correspondingly larger
% vertical spacing.  Please do not double-space your paper.  TrueType or
% Postscript Type 1 fonts are preferred.

% The first paragraph in each section should not be indented, but all the
% following paragraphs within the section should be indented as these paragraphs
% demonstrate.

\section{Future work}
This work is motivated by the fact that, disfluency detection is not only useful for the intelligent agents but also a practical problem definition to help users to produce a better, confident and fluent talk. To the extent of the types of disfluencies produced in a speech, this work is a small step towards a bigger goal, repairing disfluencies in a speech from a speakers' perspective. Along with the pitfalls of our method following could be the future directions of this work-
\begin{itemize}
\item Improving the filler word segmentation performance as well as devising techniques to segment other kinds of common disfluencies (repetition, discourse markers, corrections) and speech impairments (stuttering).
\item Devising a dynamic and online repair scheme, by generating necessary (\textit{disfluent}) portions of speech, instead of replacing.
\end{itemize}

\section{Conclusion}
Disfluency detection is a well-explored problem in the speech processing community and performed on speech transcripts to mostly aid the intelligent conversational agents. In this work, we interpret disfluency detection from speakers perspective and introduce an additional component of repairing the disfluencies. Consequently, we tried to work solely on the acoustic domain, diminishing a need for a complex system like an ASR, before disfluency detection. With the results of our detection and repair scheme, we show improved fluency in speakers' dialogues, given a less-fluent speech. To the best of our knowledge, this is the first work related to disfluency repair for the sake of users' and can be further extended to assist users with speech impairments and other general disfluencies.   

\section{Acknowledgements}
We are thankful to Faizaan Charania and Mahima Parashar for curating the dataset and working on some essential observations. We would also like to thank the participating speakers for the speeches they provided. We gratefully acknowledge the support of NVIDIA Corporation with the donation of the Titan Xp and P6000 GPU used for this research.
\balance
\bibliographystyle{IEEEbib}
\bibliography{refs}

\begin{thebibliography}{10}

\bibitem{tree1995effects}
Jean E~Fox Tree,
\newblock ``The effects of false starts and repetitions on the processing of
  subsequent words in spontaneous speech,''
\newblock {\em Journal of memory and language}, vol. 34, no. 6, pp. 709--738,
  1995.

\bibitem{womack2012disfluencies}
Kathryn Womack, Wilson McCoy, Cecilia~Ovesdotter Alm, Cara Calvelli, Jeff~B
  Pelz, Pengcheng Shi, and Anne Haake,
\newblock ``Disfluencies as extra-propositional indicators of cognitive
  processing,''
\newblock in {\em Proceedings of the workshop on extra-propositional aspects of
  meaning in computational linguistics}. Association for Computational
  Linguistics, 2012, pp. 1--9.

\bibitem{corley2008hesitation}
Martin Corley and Oliver~W Stewart,
\newblock ``Hesitation disfluencies in spontaneous speech: The meaning of um,''
\newblock {\em Language and Linguistics Compass}, vol. 2, no. 4, pp. 589--602,
  2008.

\bibitem{barr2010role}
Dale~J Barr and Mandana Seyfeddinipur,
\newblock ``The role of fillers in listener attributions for speaker
  disfluency,''
\newblock {\em Language and Cognitive Processes}, vol. 25, no. 4, pp. 441--455,
  2010.

\bibitem{shriberg1996disfluencies}
Elizabeth Shriberg,
\newblock ``Disfluencies in switchboard,''
\newblock in {\em Proceedings of International Conference on Spoken Language
  Processing}, 1996, vol.~96, pp. 11--14.

\bibitem{merlo2004descriptive}
Sandra Merlo and Let{\i}cia~Lessa Mansur,
\newblock ``Descriptive discourse: topic familiarity and disfluencies,''
\newblock {\em Journal of Communication Disorders}, vol. 37, no. 6, pp.
  489--503, 2004.

\bibitem{lennon1990investigating}
Paul Lennon,
\newblock ``Investigating fluency in efl: A quantitative approach,''
\newblock {\em Language learning}, vol. 40, no. 3, pp. 387--417, 1990.

\bibitem{cakir2017convolutional}
Emre Cak{\i}r, Giambattista Parascandolo, Toni Heittola, Heikki Huttunen, and
  Tuomas Virtanen,
\newblock ``Convolutional recurrent neural networks for polyphonic sound event
  detection,''
\newblock {\em arXiv preprint arXiv:1702.06286}, 2017.

\bibitem{kormos2004exploring}
Judit Kormos and Mariann D{\'e}nes,
\newblock ``Exploring measures and perceptions of fluency in the speech of
  second language learners,''
\newblock {\em System}, vol. 32, no. 2, pp. 145--164, 2004.

\bibitem{charniak2001edit}
Eugene Charniak and Mark Johnson,
\newblock ``Edit detection and parsing for transcribed speech,''
\newblock in {\em Proceedings of the second meeting of the North American
  Chapter of the Association for Computational Linguistics on Language
  technologies}. Association for Computational Linguistics, 2001, pp. 1--9.

\bibitem{honal2003correction}
Matthias Honal and Tanja Schultz,
\newblock ``Correction of disfluencies in spontaneous speech using a
  noisy-channel approach,''
\newblock in {\em Eighth European Conference on Speech Communication and
  Technology}, 2003.

\bibitem{johnson2004tag}
Mark Johnson and Eugene Charniak,
\newblock ``A tag-based noisy-channel model of speech repairs,''
\newblock in {\em Proceedings of the 42nd Annual Meeting of the Association for
  Computational Linguistics (ACL-04)}, 2004.

\bibitem{zwarts2010detecting}
Simon Zwarts, Mark Johnson, and Robert Dale,
\newblock ``Detecting speech repairs incrementally using a noisy channel
  approach,''
\newblock in {\em Proceedings of the 23rd International Conference on
  Computational Linguistics}. Association for Computational Linguistics, 2010,
  pp. 1371--1378.

\bibitem{liu2006enriching}
Yang Liu, Elizabeth Shriberg, Andreas Stolcke, Dustin Hillard, Mari Ostendorf,
  and Mary Harper,
\newblock ``Enriching speech recognition with automatic detection of sentence
  boundaries and disfluencies,''
\newblock {\em IEEE Transactions on audio, speech, and language processing},
  vol. 14, no. 5, pp. 1526--1540, 2006.

\bibitem{georgila2009using}
Kallirroi Georgila,
\newblock ``Using integer linear programming for detecting speech
  disfluencies,''
\newblock in {\em Proceedings of Human Language Technologies: The 2009 Annual
  Conference of the North American Chapter of the Association for Computational
  Linguistics, Companion Volume: Short Papers}. Association for Computational
  Linguistics, 2009, pp. 109--112.

\bibitem{howes2012helping}
Christine Howes, Matt Purver, Rose McCabe, PG~Healey, and Mary Lavelle,
\newblock ``Helping the medicine go down: Repair and adherence in
  patient-clinician dialogues,''
\newblock in {\em Proceedings of the 16th SemDial Workshop on the Semantics and
  Pragmatics of Dialogue (SeineDial)}, 2012, pp. 19--21.

\bibitem{honnibal2014joint}
Matthew Honnibal and Mark Johnson,
\newblock ``Joint incremental disfluency detection and dependency parsing,''
\newblock {\em Transactions of the Association of Computational Linguistics},
  vol. 2, no. 1, pp. 131--142, 2014.

\bibitem{hough2014strongly}
Julian Hough and Matthew Purver,
\newblock ``Strongly incremental repair detection,''
\newblock {\em arXiv preprint arXiv:1408.6788}, 2014.

\bibitem{howes2014helping}
Christine Howes, Julian Hough, Matthew Purver, and Rose McCabe,
\newblock ``Helping, i mean assessing psychiatric communication: An application
  of incremental self-repair detection,''
\newblock 2014.

\bibitem{ferguson2015disfluency}
James Ferguson, Greg Durrett, and Dan Klein,
\newblock ``Disfluency detection with a semi-markov model and prosodic
  features,''
\newblock in {\em Proceedings of the 2015 Conference of the North American
  Chapter of the Association for Computational Linguistics: Human Language
  Technologies}, 2015, pp. 257--262.

\bibitem{qian2013disfluency}
Xian Qian and Yang Liu,
\newblock ``Disfluency detection using multi-step stacked learning,''
\newblock in {\em Proceedings of the 2013 Conference of the North American
  Chapter of the Association for Computational Linguistics: Human Language
  Technologies}, 2013, pp. 820--825.

\bibitem{rasooli2013joint}
Mohammad~Sadegh Rasooli and Joel Tetreault,
\newblock ``Joint parsing and disfluency detection in linear time,''
\newblock in {\em Proceedings of the 2013 Conference on Empirical Methods in
  Natural Language Processing}, 2013, pp. 124--129.

\bibitem{hough2015recurrent}
Julian Hough and David Schlangen,
\newblock ``Recurrent neural networks for incremental disfluency detection,''
\newblock {\em Interspeech 2015}, 2015.

\bibitem{wang2016neural}
Shaolei Wang, Wanxiang Che, and Ting Liu,
\newblock ``A neural attention model for disfluency detection,''
\newblock in {\em Proceedings of COLING 2016, the 26th International Conference
  on Computational Linguistics: Technical Papers}, 2016, pp. 278--287.

\bibitem{zayats2016disfluency}
Vicky Zayats, Mari Ostendorf, and Hannaneh Hajishirzi,
\newblock ``Disfluency detection using a bidirectional lstm,''
\newblock {\em arXiv preprint arXiv:1604.03209}, 2016.

\bibitem{hough2017joint}
Julian Hough and David Schlangen,
\newblock ``Joint, incremental disfluency detection and utterance segmentation
  from speech,''
\newblock in {\em Proceedings of the Annual Meeting of the European Chapter of
  the Association for Computational Linguistics (EACL)}, 2017.

\bibitem{wang2017transition}
Shaolei Wang, Wanxiang Che, Yue Zhang, Meishan Zhang, and Ting Liu,
\newblock ``Transition-based disfluency detection using lstms,''
\newblock in {\em Proceedings of the 2017 Conference on Empirical Methods in
  Natural Language Processing}, 2017, pp. 2785--2794.

\bibitem{tanveer2016automanner}
M~Iftekhar Tanveer, Ru~Zhao, Kezhen Chen, Zoe Tiet, and Mohammed~Ehsan Hoque,
\newblock ``Automanner: An automated interface for making public speakers aware
  of their mannerisms,''
\newblock in {\em Proceedings of the 21st International Conference on
  Intelligent User Interfaces}. ACM, 2016, pp. 385--396.

\bibitem{chen2015pronunciation}
Guoguo Chen, Hainan Xu, Minhua Wu, Daniel Povey, and Sanjeev Khudanpur,
\newblock ``Pronunciation and silence probability modeling for asr,''
\newblock in {\em Sixteenth Annual Conference of the International Speech
  Communication Association}, 2015.

\bibitem{riggenbach1991toward}
Heidi Riggenbach,
\newblock ``Toward an understanding of fluency: A microanalysis of nonnative
  speaker conversations,''
\newblock {\em Discourse processes}, vol. 14, no. 4, pp. 423--441, 1991.

\bibitem{garofolo1993darpa}
John~S Garofolo, Lori~F Lamel, William~M Fisher, Jonathan~G Fiscus, and David~S
  Pallett,
\newblock ``Darpa timit acoustic-phonetic continous speech corpus cd-rom. nist
  speech disc 1-1.1,''
\newblock {\em NASA STI/Recon technical report n}, vol. 93, 1993.

\bibitem{sainath2015convolutional}
Tara~N Sainath, Oriol Vinyals, Andrew Senior, and Ha{\c{s}}im Sak,
\newblock ``Convolutional, long short-term memory, fully connected deep neural
  networks,''
\newblock in {\em Acoustics, Speech and Signal Processing (ICASSP), 2015 IEEE
  International Conference on}. IEEE, 2015, pp. 4580--4584.

\bibitem{cho2014learning}
Kyunghyun Cho, Bart Van~Merri{\"e}nboer, Caglar Gulcehre, Dzmitry Bahdanau,
  Fethi Bougares, Holger Schwenk, and Yoshua Bengio,
\newblock ``Learning phrase representations using rnn encoder-decoder for
  statistical machine translation,''
\newblock {\em arXiv preprint arXiv:1406.1078}, 2014.

\bibitem{rafii2012music}
Zafar Rafii and Bryan Pardo,
\newblock ``Music/voice separation using the similarity matrix.,''
\newblock in {\em ISMIR}, 2012, pp. 583--588.

\bibitem{godfrey1992switchboard}
John~J Godfrey, Edward~C Holliman, and Jane McDaniel,
\newblock ``Switchboard: Telephone speech corpus for research and
  development,''
\newblock in {\em Acoustics, Speech, and Signal Processing, 1992. ICASSP-92.,
  1992 IEEE International Conference on}. IEEE, 1992, vol.~1, pp. 517--520.

\end{thebibliography}

\end{document}